\documentclass[twocolumn]{revtex4-1}
\usepackage{amsmath,graphicx}

\begin{document}

\title{Multi-Stream Inflation: Bifurcations and Recombinations in the Multiverse}

\author{Yi Wang}
\email{wangyi@hep.physics.mcgill.ca}

\affiliation{Physics Department, McGill University, Montreal, H3A2T8, Canada}


\begin{abstract}

In this Letter, we briefly review the multi-stream inflation scenario, and discuss its implications in the string theory landscape and the inflationary multiverse. In multi-stream inflation, the inflation trajectory encounters bifurcations. If these bifurcations are in the observable stage of inflation, then interesting observational effects can take place, such as domain fences, non-Gaussianities, features and asymmetries in the CMB. On the other hand, if the bifurcation takes place in the eternal stage of inflation, it provides an alternative creation mechanism of bubbles universes in eternal inflation, as well as a mechanism to locally terminate eternal inflation, which reduces the measure of eternal inflation.

\end{abstract}
\maketitle

\section{Introduction}

Inflation \cite{Guth:1980zm} has become the leading paradigm for the very early universe. However, the detailed mechanism for inflation still remains unknown. Inspired by the picture of string theory landscape \cite{Bousso:2000xa}, one could expect that the inflationary potential has very complicated structure \cite{Huang:2008jr}. Inflation in the string theory landscape has important implications in both observable stage of inflation and eternal inflation.

The complicated inflationary potentials in the string theory landscape open up a great number of interesting observational effects during observable inflation. Researches investigating the complicated structure of the inflationary potential include multi-stream inflation \cite{Li:2009sp, Li:2009me}, quasi-single field inflation \cite{Chen:2009we}, meandering inflation \cite{Tye:2009ff}, old curvaton \cite{Gong:2008ni}, etc.

The string theory landscape also provides a playground for eternal inflation. Eternal inflation is an very early stage of inflation, during which the universe reproduces itself, so that inflation becomes eternal to the future. Eternal inflation, if indeed happened (for counter arguments see, for example \cite{Mukhanov:1996ak}), can populate the string theory landscape, providing an explanation for the cosmological constant problem in our bubble universe by anthropic arguments.

\begin{figure}
\includegraphics[width=0.37\textwidth]{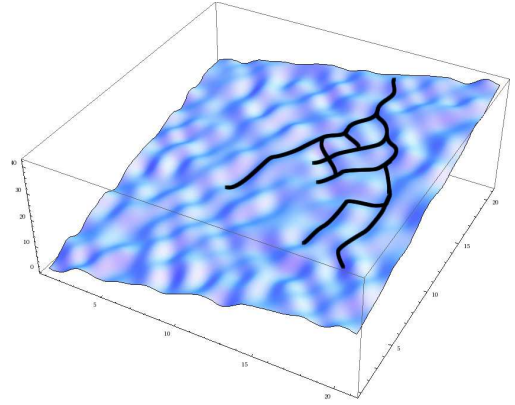}
\caption{In this figure, we use a tilted random potential to mimic a inflationary potential in the string theory landscape. One can expect that in such a random potential, bifurcation effects happens generically, as illustrated in the trajectories in the figure.}
\label{fig:random}
\end{figure}

In this Letter, we shall focus on the multi-stream inflation scenario. Multi-stream inflation is proposed in \cite{Li:2009sp}. And in \cite{Li:2009me}, it is pointed out that the bifurcations can lead to multiverse. Multi-stream inflation assumes that during inflation there exist bifurcation(s) in the inflation trajectory. For example, the bifurcations take place naturally in a random potential, as illustrated in Fig. \ref{fig:random}. We briefly review multi-stream inflation in Section \ref{sec:observable}. The details of some contents in Section \ref{sec:observable} can be found in \cite{Li:2009sp}. We discuss some new implications of multi-stream inflation for the inflationary multiverse in Section \ref{sec:eternal}.

\begin{figure}
\includegraphics[width=0.37\textwidth]{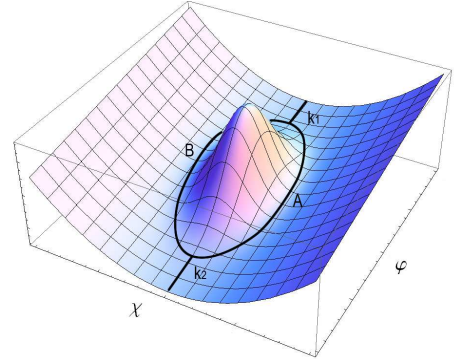}
\caption{One sample bifurcation in multi-stream inflation. The inflation trajectory bifurcates into $A$ and $B$ when the comoving scale $k_1$ exits the horizon, and recombines when the comoving scale $k_2$ exits the horizon.}
\label{fig:msinf}
\end{figure}

\section{Observable bifurcations} \label{sec:observable}
In this section, we discuss the possibility that the bifurcation of multi-stream inflation happens during the observable stage of inflation.  We review the production of non-Gaussianities, features and asymmetries \cite{Li:2009sp} in the CMB, and investigate some other possible observational effects.

\begin{figure}
\includegraphics[width=0.37\textwidth]{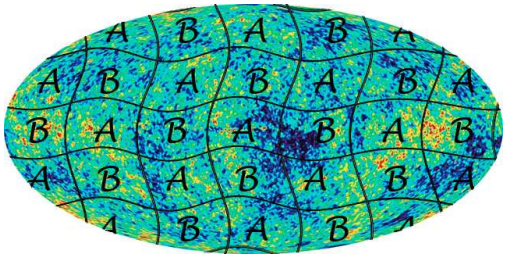}
\caption{In multi-stream inflation, the universe breaks up into patches with comoving scale $k_1$. Each patch experienced inflation either along trajectories $A$ or $B$. These different patches can be responsible for the asymmetries in the CMB.}
\label{fig:msicmb}
\end{figure}

To be explicit, we focus on one single bifurcation, as illustrated in Fig. \ref{fig:msinf}. We denote the initial (before bifurcation) inflationary direction by $\varphi$, and the initial isocurvature direction by $\chi$. For simplicity, we let $\chi=0$ before bifurcation. When comoving wave number $k_1$ exits the horizon, the inflation trajectory bifurcates into $A$ and $B$. When comoving wave number $k_2$ exits the horizon, the trajectories recombines into a single trajectory. The universe breaks into of order $k_1/k_0$ patches (where $k_0$ denotes the comoving scale of the current observable universe), each patch experienced inflation either along trajectories $A$ or $B$. The choice of the trajectories is made by the isocurvature perturbation $\delta\chi$ at scale $k_1$. This picture is illustrated in Fig. \ref{fig:msicmb}.

We shall classify the bifurcation into three cases:

{\it Symmetric bifurcation}. If the bifurcation is symmetric, in other words, $V(\varphi,\chi)=V(\varphi, -\chi)$, then there are two potentially observable effects, namely, quasi-single field inflation, and a effect from a domain-wall-like objects, which we call domain fences.

As discussed in \cite{Li:2009sp}, the discussion of the bifurcation effect becomes simpler when the isocurvature direction has mass of order the Hubble parameter. In this case, except for the bifurcation and recombination points, trajectory $A$ and trajectory $B$ experience quasi-single field inflation respectively. As there are turnings of these trajectories, the analysis in \cite{Chen:2009we} can be applied here. The perturbations, especially non-Gaussianities in the isocurvature directions are projected onto the curvature direction, resulting in a correction to the power spectrum, and potentially large non-Gaussianities. As shown in \cite{Chen:2009we}, the amount of non-Gaussianity is of order
\begin{equation}
  f_{NL} \sim P_\zeta^{-1/2} \left(\frac{1}{H}\frac{\partial^3 V}{\partial \chi^3}\right) \left(\frac{\dot\theta}{H}\right)^3~,
\end{equation}
where $\theta$ denotes the angle between the true inflation direction and the $\varphi$ direction.

As shown in Fig. \ref{fig:msicmb}, the universe is broken into patches during multi-stream inflation. There are wall-like boundaries between these patches. During inflation, these boundaries are initially domain walls. However, after the recombination of the trajectories, the tensions of these domain walls vanish. We call these objects domain fences. As is well known, domain wall causes disasters in cosmology because of its tension. However, without tension, domain fence does not necessarily cause such disasters. It is interesting to investigate whether there are observational sequences of these domain fences.

{\it Nearly symmetric bifurcation}
If the bifurcation is nearly symmetric, in other words, $V(\varphi,\chi) \simeq V(\varphi, -\chi)$, but not equal exactly, which can be achieved by a spontaneous breaking and restoring of an approximate symmetry, then besides the quasi-single field effect and the domain fence effect, there will be four more potentially observable effects in multi-stream inflation, namely, the features and asymmetries in CMB, non-Gaussianity at scale $k_1$ and squeezed non-Gaussianity correlating scale $k_1$ and scale $k$ with $k_1<k<k_2$.

The CMB power asymmetries are produced because, as in Fig. \ref{fig:msicmb}, patches coming from trajectory $A$ or $B$ can have different power spectra $P_\zeta^A$ and $P_\zeta^B$, which are determined by their local potentials. If the scale $k_1$ is near to the scale of the observational universe $k_0$, then multi-stream inflation provides an explanation of the hemispherical asymmetry problem \cite{Erickcek:2008sm}.

The features in the CMB (here feature denotes extra large perturbation at a single scale $k_1$) are produced as a result of the e-folding number difference $\delta N$ between two trajectories. From the $\delta N$ formalism, the curvature perturbation in the uniform density slice at scale $k_1$ has an additional contribution
\begin{equation}
  \delta\zeta_{k_1}\sim \delta N \equiv |N_A-N_B|~.
\end{equation}
These features in the CMB are potentially observable in the future precise CMB measurements. As the additional fluctuation $\delta\zeta_{k_1}$ does not obey Gaussian distribution, there will be non-Gaussianity at scale $k_1$.

Finally, there are also correlations between scale $k_1$ and scale $k$ with $k_1<k<k_2$. This is because the additional fluctuation $\delta\zeta_{k_1}$ and the asymmetry at scale $k$ are both controlled by the isocurvature perturbation at scale $k_1$. Thus the fluctuations at these two scales are correlated. As estimated in \cite{Li:2009sp}, this correlation results in a non-Gaussianity of order
\begin{equation}
  f_{NL}\sim \frac{\delta\zeta_{k_1}}{\zeta_{k_1}} \frac{P_\zeta^A-P_\zeta^B}{P_\zeta^A} P_\zeta^{-1/2}~.
\end{equation}

{\it Non-symmetric bifurcation}
If the bifurcation is not symmetric at all, especially with large e-folding number differences (of order ${\cal O}(1)$ or greater) along different trajectories, the anisotropy in the CMB and the large scale structure becomes too large at scale $k_1$. However, in this case, regions with smaller e-folding number will have exponentially small volume compared with regions with larger e-folding number. Thus the anisotropy can behave in the form of great voids. We shall address this issue in more detail in \cite{prep}. Trajectories with e-folding number difference from ${\cal O}(10^{-5})$ to ${\cal O}(1)$ in the observable stage of inflation are ruled out by the large scale isotropy of the observable universe.

At the remainder of this section, we would like to make several additional comments for multi-stream inflation:

{\it The possibility that the bifurcated trajectories never recombine}. In this case, one needs to worry about the domain walls, which do not become domain fence during inflation. These domain walls may eventually become domain fence after reheating anyway. Another problem is that the e-folding numbers along different trajectories may differ too much, which produce too much anisotropies in the CMB and the large scale structure. However, similar to the discussion in the case of non-symmetric bifurcation, in this case, the observable effect could become great voids due to a large e-folding number difference. The case without recombination of trajectory also has applications in eternal inflation, as we shall discuss in the next section.

{\it Probabilities for different trajectories}. In \cite{Li:2009sp}, we considered the simple example that during the bifurcation, the inflaton will run into trajectories $A$ and $B$ with equal probabilities. Actually, this assumption does not need to be satisfied for more general cases. The probability to run into different trajectories can be of the same order of magnitude, or different exponentially. In the latter case, there is a potential barrier in front of one trajectory, which can be leaped over by a large fluctuation of the isocurvature field. A large fluctuation of the isocurvature field is exponentially rare, resulting in exponentially different probabilities for different trajectories.  The bifurcation of this kind is typically non-symmetric.

{\it Bifurcation point itself does not result in eternal inflation}. As is well known, in single field inflation, if the inflaton releases at a local maxima on a ``top of the hill'', a stage of eternal inflation is usually obtained. However, at the bifurcation point, it is not the case. Because although the $\chi$ direction releases at a local maxima, the $\varphi$ direction keeps on rolling at the same time. The inflation direction is a combination of these two directions. So multi-stream inflation can coexist with eternal inflation, but itself is not necessarily eternal.

\section{Eternal bifurcations} \label{sec:eternal}

In multi-stream inflation, the bifurcation effect may either take place at an eternal stage of inflation. In this case, it provides interesting ingredients to eternal inflation. These ingredients include alternative mechanism to produce different bubble universes and local terminations for eternal inflation, as we shall discuss separately.

{\it Multi-stream bubble universes}. The most discussed mechanisms to produce bubble universes are tunneling processes, such as Coleman de Luccia instantons \cite{Coleman:1980aw} and Hawking Moss instantons \cite{Hawking:1981fz}. In these processes, the tunneling events, which are usually exponentially suppressed, create new bubble universes, while most parts of the spatial volume remain in the old bubble universe at the instant of tunneling.

If bifurcations of multi-stream inflation happen during eternal inflation, two kinds of new bubble universes can be created with similar probabilities. In this case, at the instant of bifurcation, both kinds of bubble universes have nearly equal spatial volume. With a change of probabilities, the measures for eternal inflation should be reconsidered for multi-stream type bubble creation mechanism.

If the inflation trajectories recombine after a period of inflation, the different bubble universes will eventually have the same physical laws and constants of nature. On the other hand, if the different inflation trajectories do not recombine, then the different bubble universes created by the bifurcation will have different vacuum expectation values of the scalar fields, resulting to different physical laws or constants of nature. It is interesting to investigate whether the bifurcation effect is more effective than the tunneling effect to populate the string theory landscape.

Note that in multi-stream inflation, it is still possible that different trajectories have exponentially different probabilities, as discussed in the previous section. In this case, multi-stream inflation behaves similar to Hawking Moss instantons during eternal inflation.

\begin{figure}
\includegraphics[width=0.37\textwidth]{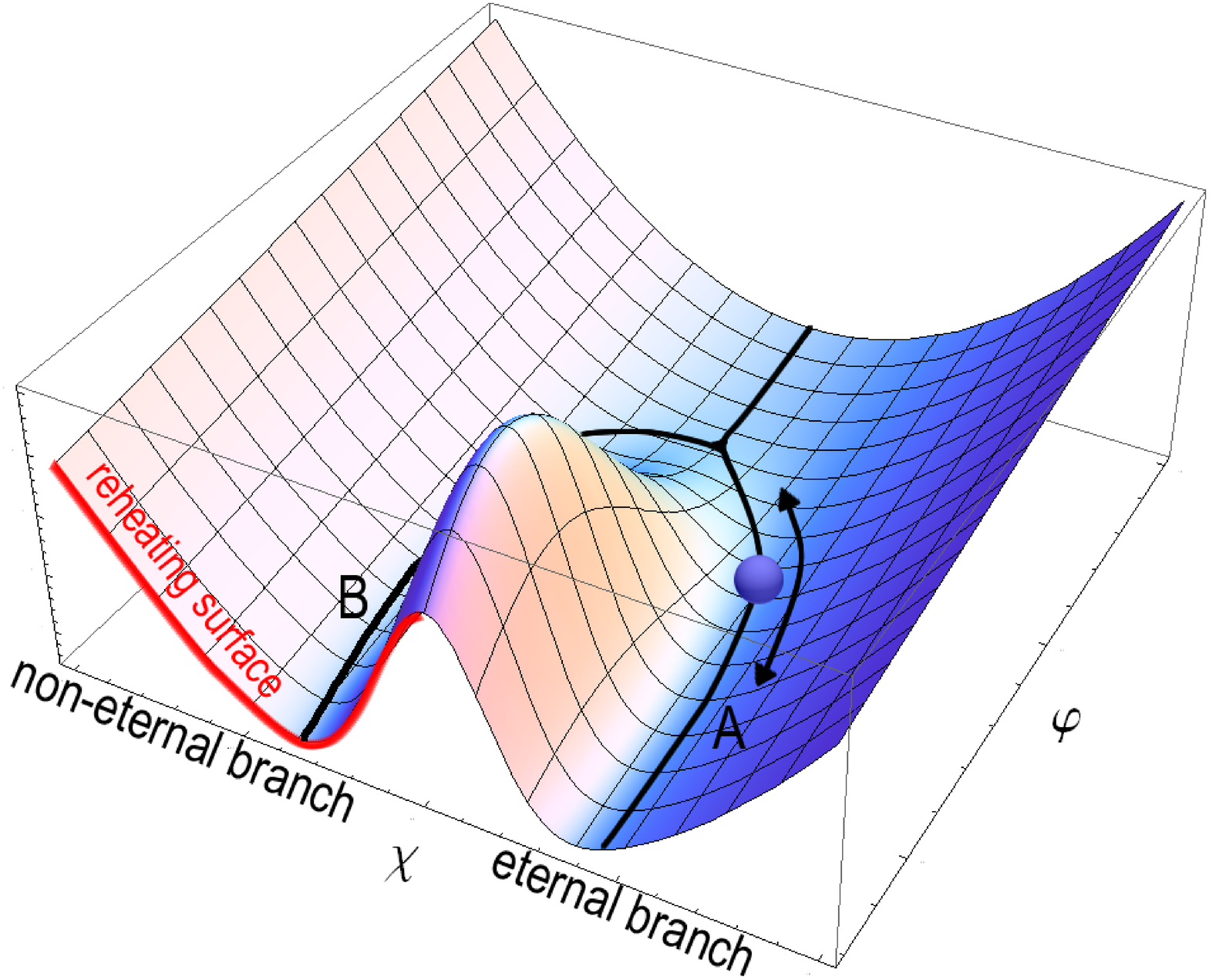}

\vspace{2mm}
\includegraphics[width=0.37\textwidth]{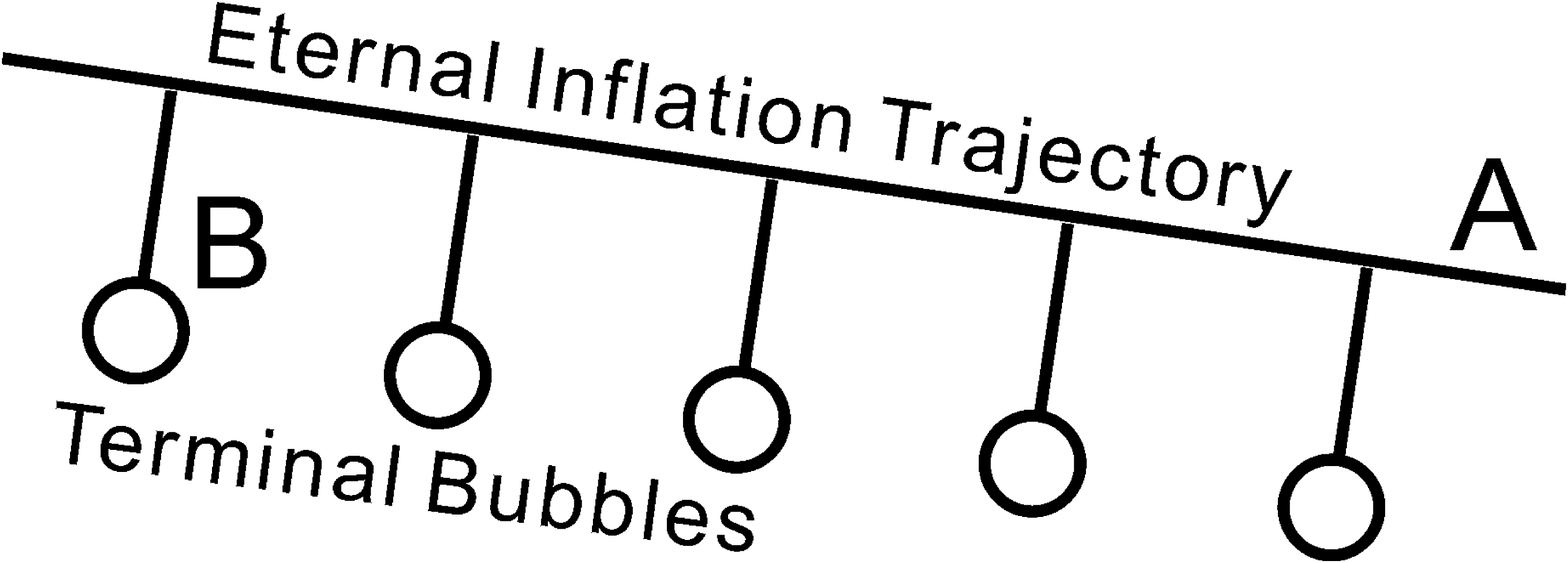}
\caption{Cascade creation of bubble universes. In this figure, we assume trajectory $A$ is the eternal inflation trajectory, and trajectory $B$ is the non-eternal inflation trajectory.}
\label{fig:cascade}
\end{figure}

{\it Local terminations for eternal inflation}. It is possible that during multi-stream inflation, a inflation trajectory bifurcates in to one eternal inflation trajectory and one non-eternal inflation trajectory with similar probability. In this case, the inflaton in the eternal inflation trajectory frequently jumps back to the bifurcation point, resulting in a cascade creation of bubble universes, as illustrated in Fig. \ref{fig:cascade}. This cascade creation of bubble universes, if realized, is more efficient in producing reheating bubbles than tunneling effects. Thus it reduces the measure for eternal inflation.

There are some other interesting issues for bifurcation in the multiverse. For example, the bubble walls may be observable in the present observable universe, and the bifurcations can lead to multiverse without eternal inflation. These possibilities are discussed in \cite{Li:2009me}.

\section{Conclusion and discussion}

To conclude, we briefly reviewed multi-stream inflation during observable inflation. Some new issues such as domain fences and connection with quasi-single field inflation are discussed. We also discussed multi-stream inflation in the context of eternal inflation. The bifurcation effect in multi-stream inflation provides an alternative mechanism for creating bubble universes and populating the string theory landscape. The bifurcation effect also provides a very efficient mechanism to locally terminate eternal inflation.

\section*{Acknowledgment}
We thank Yifu Cai for discussion. This work was supported by NSERC and an IPP postdoctoral fellowship.

\end{document}